\documentclass{article}

    \PassOptionsToPackage{numbers, compress}{natbib}
 \usepackage[preprint]{neurips_2026}

\usepackage[utf8]{inputenc} 
\usepackage[T1]{fontenc}    
\usepackage{hyperref}       
\usepackage{url}            
\usepackage{booktabs}       
\usepackage{amsfonts}       
\usepackage{nicefrac}       
\usepackage{microtype}      
\usepackage{xcolor}         
\usepackage{amsmath}
\usepackage[pdftex]{graphicx}
\usepackage{tikz}
\usetikzlibrary{positioning, arrows.meta, fit, backgrounds, shapes.geometric, decorations.pathreplacing}
\usepackage{longtable}
\usepackage{array}

\title{MIRAI: Prediction and Generation of High-Impact Academic Research}

\author{%
  Alex Li \\
  MIT Media Lab\\
  Cambridge, MA\\
  \texttt{alexzli@mit.edu} \\
  \And
  Joseph Jacobson \\
  MIT Media Lab \\
  Cambridge, MA \\
  \texttt{jacobson@media.mit.edu} \\
}

\begin{document}

\maketitle

\begin{abstract}
  The rapid pace of scientific publishing has made the identification and synthesis of high-impact work an increasingly urgent challenge. We introduce MIRAI (Multi-year Inference of Research trends and Academic Impact), a deep learning framework that predicts paper impact using only it's title, abstract, and publication date. We train MIRAI on the arXiv academic graph to predict 5-year PageRank and citation counts, achieving Spearman's $\rho$ of 0.4686 on PageRank prediction and 0.6192 on citation prediction for papers published in 2021. We propose a research ideation pipeline built on top of MIRAI that produces research ideas oriented towards high impact. These ideas were judged as more impactful than a baseline without MIRAI by an unbiased LLM judge at a 4:3 ratio. We make the 5-year citation prediction model publicly available at \url{https://predict-paper-impact.vercel.app}.
\end{abstract}

\section{Introduction}
The volume of scientific literature has entered an era of near-exponential growth \cite{10.1162/qss_a_00327}. In 2024, the Nature Index recorded its largest single-year increase in primary research articles across the natural and health sciences, rising 16.1\% over 2023 \cite{baker2025natureindex}. The trend is equally stark on arXiv: March 2026 saw a record 30,045 submissions in a single month, pushing the platform past 3 million total articles, a milestone reached only four years after arXiv hit 2 million, compared to the eight years it took to grow from 1 to 2 million \cite{arxiv2026stats}. 

Navigating this ever-growing corpus of scientific literature is becoming an increasingly difficult task for researchers, funding bodies, and institutions attempting to allocate intellectual and financial resources effectively. The challenge is compounded by a growing proportion of low-quality papers, enabled by the widespread availability of public datasets and the rapid proliferation of large language models (LLMs) that facilitate AI-generated text production at scale \cite{ogrady2025lowquality,suchak2025nhanes}. As the volume of published research continues to rise — and an increasing share of it is AI-generated — the ability to efficiently and reliably identify high-value work from within the broader literature becomes not merely useful, but essential.

These pressures on the output side of science are mirrored by pressures on the input side. In recent years, research funding in the US has contracted sharply. In 2025, the NSF issued 25\% fewer new grants than its ten-year average, while the NIH issued 24\% fewer, reductions accompanied by substantial staff cuts across federal science agencies \cite{kozlov2026trump}. These pressures arrive at a moment when the peer review process that governs research funding allocation is itself facing growing scrutiny. Critics have pointed to its reliance on crude quantitative proxies, such as publication count or journal impact factor, and to evidence of unconscious bias systematically disadvantaging certain applicants \cite{bendiscioli2019peerreview}. Together, these forces make the question of how to identify and prioritize high-impact research both more urgent and more consequential.

While the rise of LLMs has exacerbated the problem of exploding research volume, it may also be key to navigating it. Language models have already demonstrated considerable promise across tasks closely related to research evaluation, including automated peer review \cite{liang2024llmreview, darcy2024marg}, citation impact prediction directly from paper abstracts and titles \cite{vital2024predictingcitationimpactresearch, zhao2024wordsworthnewbornarticle}, and novel hypothesis generation \cite{si2024llmideas}. Existing approaches to identifying high-impact research largely rely on peer review, which, in addition to the scalability and bias concerns discussed above, cannot keep pace with the current rate of publication. Quantitative alternatives typically depend on citation-based metrics, which are inherently lagging indicators, often requiring years after publication to fully materialize and are prone to amplifying author and journal prestige effects. Analyzing research impact directly from content using language models is both lower-latency and less susceptible to such confounds, and could represent a more scalable and equitable basis for evaluating the merit of scientific work.

This work pursues two related questions that emerge from these challenges. The first is whether research impact can be predicted at the paper level from content available at publication time, providing a scalable, low-latency signal for literature filtering, recommendation, and resource allocation. The second is whether that predictive signal can be turned generative: if a model can learn what makes research high-impact, can it also help synthesize novel research directions that are likely to be?

\subsection{Approach and Contributions}

The rise of low-latency scientific publishing, particularly arXiv preprints, has transformed how research is disseminated, making it possible to observe and evaluate new work months before it enters the traditional citation ecosystem \cite{aman2013potentialpreprintsacceleratescholarly,xie2021preprintfuturesciencethirty}. Unlike traditionally published papers, which may take months or years to appear in indexed databases and begin accumulating citations, arXiv preprints are publicly available immediately upon submission, representing the true frontier of scientific output. We therefore restrict our scope entirely to arXiv papers, constructing an academic citation graph exclusively from this corpus using the Semantic Scholar API \cite{Kinney2023TheSS}. However, we acknowledge that the arXiv corpus is overrepresented by papers in computer science, mathematics, and physics. We leave generalizing our work to other fields and journals as future work.

We propose a MIRAI, Multi-year Inference of Research trends and Academic Impact, a machine learning framework for predicting the scientific impact of academic papers using only a universal text embedding of a paper's title and abstract as input. Using our model, we achieve Spearman's $\rho$ of 0.62 on 5-year citation prediction and 0.47 on 5-year PageRank prediction on papers published in 2021.

We then propose a research generation pipeline that uses our academic graph and trained prediction model to guide the ideation process, producing novel titles and abstracts oriented toward high-impact research directions. 

In summary, this work makes the following contributions:
\begin{itemize}
    \item \textbf{Dataset:} A dataset of nearly 3 million arXiv papers with authorship, citation, and network-based impact labels, including both citation counts and PageRank scores.
    \item \textbf{Impact prediction:} A machine learning framework for predicting different measures of paper impact using a paper's title, abstract, and publication date.
    \item \textbf{Research generation:} A research generation pipeline that leverages our novel impact prediction framework.
\end{itemize}

We release an impact prediction model trained on 5-year citation outcomes at \url{https://predict-paper-impact.vercel.app}.

\section{Related Work}

We situate our work within two bodies of related research that have largely been studied in isolation: impact prediction from scientific text, and AI-assisted generation of scientific content. We begin with an overview of measures of scientific impact.

\subsection{Measures of Impact}

The most direct way of assessing scientific impact is through citation-based metrics such as citation counts or the h-index. However, such metrics have long been criticized for their susceptibility to bias and manipulation, their field-dependence, and their tendency to disadvantage both recently published work and early-career researchers who have not yet had time to accumulate citations \cite{nature2005impactfactor,doi:10.1126/science.1212540,doi:10.1126/science.aaa3796}.

Network-based alternatives such as PageRank measure a paper's centrality in the citation graph by weighting citations from highly-cited papers more heavily than those from less-cited ones \cite{Page1998PageRank}. These metrics have been shown to better identify milestone papers, particularly when scaled for publication age \cite{Mariani_2016,Xu_2020}. We therefore adopt PageRank as our primary measure of long-term scientific influence. We additionally report citation count for two reasons: it is the dominant target metric in prior impact prediction work, enabling direct comparison with existing models, and it is more readily interpretable than PageRank. 

\subsection{Research Impact Prediction}

The prediction of research impact has been an active area of study for well over a decade. Early work focused on predicting citation counts using hand-crafted features drawn from a paper's content, authorship, and publication venue, fed into regression models, SVMs, or other classical machine learning methods \cite{fu2008models,10.1145/2063576.2063757}. More recent work has refined this paradigm: Li \textit{et al.} trained a multi-layer perceptron on hand-crafted features for biomedical papers, augmenting them with citation information gathered two years post-publication \cite{li2022predictingclinicalcitationcount}. Weis and Jacobson take a different target metric entirely, using a tree-based model on crafted features to predict a time-rescaled PageRank score rather than raw citation count \cite{weis2021delphi}, a measure that has been shown to better capture long-term research influence. However, a common limitation of these two approaches is their dependence on post-publication data, with both requiring citation or network information gathered years after publication. Additionally, neither makes full use of a paper's textual content: Weis and Jacobson incorporate none at all, while Li \textit{et al.} include only shallow, domain-specific content features that limit generalizability beyond the biomedical literature.

Recent advances in language models have opened new avenues for text-driven impact prediction through dense text embeddings. Vital Jr. \textit{et al.} frame the problem as binary classification, predicting whether a paper will fall above or below a citation count percentile cutoff, and test \texttt{text-embedding-ada-002} embeddings alongside other embedding techniques across multiple model architectures \cite{vital2024predictingcitationimpactresearch}. Notably, they achieve 80\% classification accuracy using only titles and abstracts, with no authorship, venue, or citation information. Hirako \textit{et al.} use a BERT-based embedding to perform regression on citation counts for papers from select categories on arXiv and bioRxiv, achieving a Spearman correlation of 0.436 among computational linguistics papers \cite{hirako2024cimatecitationcountprediction}. Together, these results suggest that deep text representations alone carry substantial signal about a paper's future influence.

A parallel line of work attempts to extract impact signals from LLMs directly. Zhao \textit{et al.} perform parameter-efficient fine-tuning (PEFT) to an LLM to predict a novel impact metric from title and abstract alone in what they call ``newborn article impact prediction'' \cite{zhao2024wordsworthnewbornarticle}. Ye \textit{et al.} take a prompt engineering approach, querying LLMs zero-shot to classify papers as high or low impact based on citation counts \cite{ye2026largelanguagemodelsable}. While these approaches yield compelling results, directly querying LLMs at inference time is significantly slower and more computationally expensive than embedding-based prediction, and existing results suggest little performance gain over simpler embedding-based methods despite the additional compute cost. We therefore chose to pursue an embedding-based approach for our work.

\subsection{Scientific Content Generation}

LLMs have rapidly emerged as powerful tools for accelerating scientific discovery, being used at every stage of the scientific process from literature analysis and hypothesis formulation to experimental design and result interpretation \cite{zheng2025automationautonomysurveylarge}. More recently, LLMs have moved beyond being tools for human researchers, towards acting as autonomous agents capable of independent scientific discovery and even end-to-end paper generation \cite{lu2024aiscientistfullyautomated}.

A particularly relevant development for this work is the use of LLMs in scientific ideation, where they can aid scientists in generating research ideas by identifying gaps in existing literature or synthesizing prior work into novel directions. Si \textit{et al.} showed that LLMs can generate research ideas rated as more novel than those produced by human researchers, underscoring their potential as creative partners in the scientific process \cite{si2024llmideas}.

Most existing approaches to LLM-based scientific ideation rely on zero-shot prompting, in which a model is given a research topic or a set of seed papers and asked to generate novel directions. For instance, Si \textit{et al.} prompt an LLM directly with a research topic and then use pairwise comparisons to filter generated ideas by relevance, feasibility, and novelty \cite{si2024llmideas}. Lu \textit{et al.} take a similar approach in their AI Scientist, prompting an LLM to generate ideas conditioned on a set of its own prior outputs, which are then refined through chain-of-thought reasoning and self-reflection before being filtered for the most novel ideas \cite{lu2024aiscientistfullyautomated}. In both cases, idea generation is not the primary contribution of the work, and no explicit objective is used to steer the model toward any particular quality of output.

More structured approaches to LLM-guided idea generation have also been developed. Li \textit{et al.} introduce a Chain-of-Ideas agent that organizes existing literature into a citation chain, using the evolution of a research domain to ground idea generation in current trends and developments \cite{li2024chainideasrevolutionizingresearch}. Zhao \textit{et al.}, in their Deep Ideation framework, analyze keyword overlap between scientific papers to map contextual relationships across research domains, providing a richer foundation for LLM-guided ideation than raw text alone \cite{zhao2025deepideationdesigningllm}. Both approaches leverage the structure of existing literature to guide the generation of ideas that are more novel and relevant than what zero-shot prompting alone can achieve. Despite these advancements, the generation of research ideas explicitly optimized for downstream academic impact remains largely unexplored and is the focus of this work.

\section{Data Collection}
\label{ch:graph}

\subsection{arXiv Graph Construction}

Our academic graph includes arXiv papers submitted up to April 9, 2026. We first obtain the full set of arXiv paper IDs from the publicly available Kaggle arXiv dataset \cite{arxiv_org_submitters_2024}. These IDs are then used to query the Semantic Scholar paper batch endpoint \cite{Kinney2023TheSS}, from which we retrieve the fields detailed in \autoref{tab:fields}. 

\begin{table}
\centering
\caption{Fields extracted from the Semantic Scholar API for each paper.}
\label{tab:fields}
\begin{tabular}{ll}
\toprule
\textbf{Field} & \textbf{Description} \\
\midrule
\texttt{paperId}               & Semantic Scholar paper identifier \\
\texttt{externalIds}           & External identifiers (e.g., arXiv ID) \\
\texttt{publicationDate}       & Date of publication \\
\texttt{title}                 & Paper title \\
\texttt{abstract}              & Paper abstract \\
\texttt{references.externalIds}& External IDs of cited papers \\
\texttt{fieldsOfStudy}         & Semantic Scholar field classification \\
\bottomrule
\end{tabular}
\end{table}

The \texttt{references.externalIds} field contains external identifiers for each cited paper, including both its Semantic Scholar ID and its arXiv ID where available. We use the presence of an arXiv ID in this field to filter references to only those papers that also appear in our arXiv corpus, ensuring the citation graph remains self-contained, which is required for PageRank computation. For papers with multiple fields of study, we retain only the primary field, defined as the first entry in the \texttt{fieldsOfStudy} list as returned by the API. 

The collected papers are represented as an in-memory directed citation graph. The graph is stored as sparse matrices in CSR format using SciPy \cite{scipy}, which enables efficient computation of network features such as citation counts and PageRank, as well as fast extraction of temporal subgraphs for time-windowed analysis. To facilitate the latter, publication dates are also stored in memory alongside the graph structure, whereas titles and abstracts are retrieved from disk as needed.

\begin{figure}
    \centering
    \includegraphics[width=0.8\textwidth]{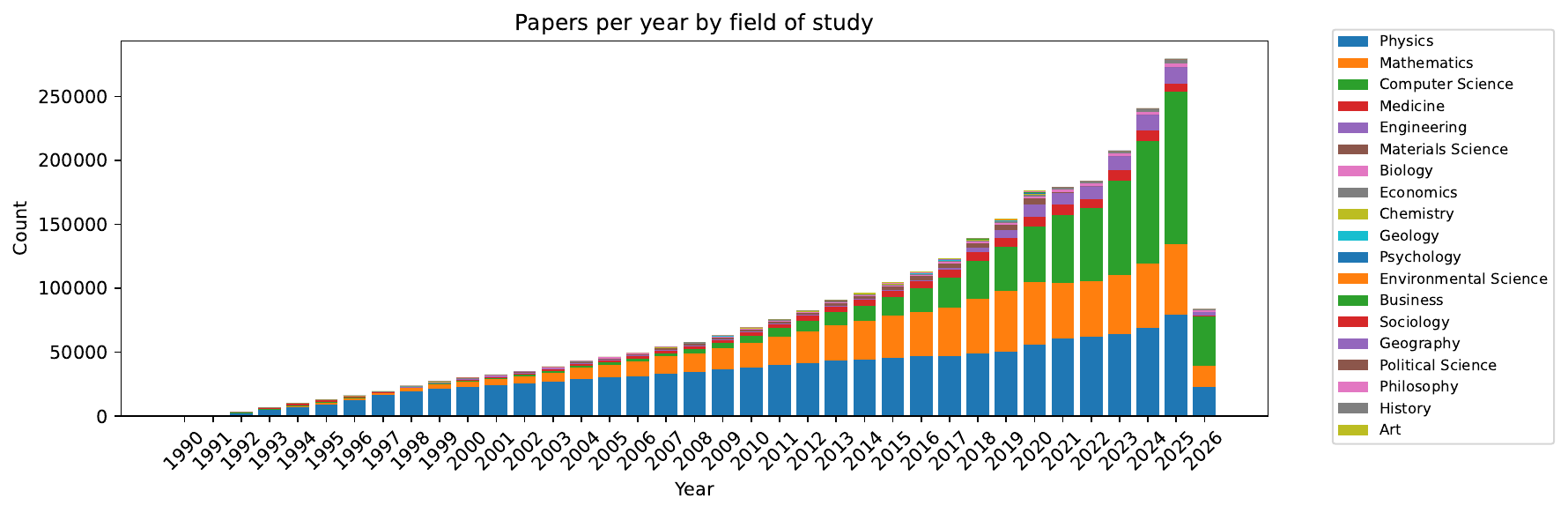}
    \caption{Number of papers published per year by field of study.}
    \label{fig:papers_per_year}
\end{figure}

The distribution of papers by year and field of study shown in \autoref{fig:papers_per_year}. The rapid growth of the number of preprints, particularly computer science papers, is evident from this graph. The dataset contains 2,976,125 papers, representing 98.74\% of the complete arXiv corpus of 3,014,049 articles at the time of writing. The earliest paper in the dataset is dated 1825, as some historical papers were submitted to arXiv after its launch in 1991.

\subsection{Labels Dataset Generation}

\subsubsection{Citation count}

Citation counts naturally accumulate over time, which means that older papers have had more opportunity to accrue citations than newer ones. To ensure a fair comparison across papers of different ages, we use the $n$-year citation count $C_n$, defined as the number of citations a paper receives within its first $n$ years of publication. Rather than using raw citation counts as prediction targets, we apply a log transformation, as citation distributions are known to be heavy-tailed \cite{Wang_2013}. We therefore define our citation score as $\ln(1 + C_n)$. Papers without a fully elapsed $n$-year window are excluded from the corresponding label set.

\paragraph{Normalization considerations.}

Typical citation-based impact metrics normalize citation counts by research field and publication year to enable direct comparisons across time frames and disciplines \cite{waltman2010newcrownindicatortheoretical}, and are often preferred over raw citation counts as measures of impact. However, we decided against both forms of normalization for several reasons. For field normalization, the fields provided by Semantic Scholar are too coarse to reliably capture the diversity of research represented in our corpus, and a single field label often fails to accurately characterize interdisciplinary work. Rather than normalizing by field, we rely on the model to learn field-relevant information directly from the title and abstract, which provides a more continuous and flexible representation of a paper's research area than a discrete field label. For year normalization, our use of a fixed citation horizon already addresses its primary motivation of ensuring that papers are compared over the same accumulation window regardless of age. While differences in citation dynamics across publication years remain a valid concern, calendar year-based normalization can produce noisy estimates for recently published papers, as noted by Ioannidis \textit{et al.} \cite{normalize}. Instead, we include publication date as an explicit input feature in our prediction models, allowing them to learn temporal citation trends directly from the data.

\subsubsection{PageRank}

Just as with citation counts, PageRank is computed at fixed time horizons to ensure comparability across papers of different ages. Papers are grouped into weekly cohorts for computational efficiency, with each cohort's PageRank computed on the citation subgraph of all papers published up to exactly $n$ calendar years after the start date of the cohort week. For example, under a 1-year horizon, papers published between January 1--7, 2024 have their PageRank evaluated on the subgraph of all papers published on or before January 1, 2025. We compute PageRank using the standard damping factor of $\alpha=0.85$, a convergence tolerance of $10^{-6}$, and a maximum of 100 power iterations \cite{Page1998PageRank}.

One challenge with raw PageRank is that for a graph of size $N$, the average PageRank score is $\frac{1}{N}$ by construction. As the citation graph grows over time, every paper's raw PageRank is diluted simply due to graph size, independent of any paper's actual impact. To correct for this, we instead compute $\ln(N \times \text{PageRank})$, where $N$ is the number of papers in the cohort's subgraph at evaluation time and thus varies per cohort. The $N \times \text{PageRank}$ quantity can be interpreted as how many times more likely a random walker is to visit a given paper than a uniformly chosen node, making it directly comparable across cohorts of different sizes. The log transformation is then applied because, as with citation counts, these scores remain heavy-tailed.

\section{Impact Prediction Methodology}

\subsection{Paper Representation}

Recent advances in language models have driven significant progress in text embedding, enabling the computation of dense vector representations that capture rich semantic content. We embed each paper's title and abstract using \texttt{llama-embed-nemotron-8b}, a text embedding model developed by NVIDIA that achieved state-of-the-art performance on the multilingual MTEB leaderboard \cite{nemotron, mteb}. The model produces 4096-dimensional dense embeddings. Title and abstract are concatenated with the format \texttt{"\{title\}\textbackslash n\{abstract\}"} before being passed to the model.

Each model takes the publication date as an additional input alongside the text embedding. The date was first converted into a fractional year and then transformed into a z-score using the mean and standard deviation of the dates from the corresponding training dataset. 

\subsection{Dataset Splitting}

\begin{table}
\centering
\caption{Label availability for each publication window in our splitting scheme. Checkmarks indicate that the $n$-year label is observable by the snapshot date for all papers in the window; dashes indicate that it is not. Each paper contributes only to the per-horizon loss terms for which its label is available.}
\label{tab:multihorizon_split}
\begin{tabular}{llccccc}
\toprule
\textbf{Publication window} & \textbf{Set} & \textbf{1y} & \textbf{2y} & \textbf{3y} & \textbf{4y} & \textbf{5y} \\
\midrule
$T{-}8$ to $T{-}6$ (2013--2015)                  & Train & \checkmark & \checkmark & \checkmark & \checkmark & \checkmark \\
Jan 1 -- Apr 9 of $T{-}5$ (2016)                 & Valid & \checkmark & \checkmark & \checkmark & \checkmark & \checkmark \\
Apr 10 of $T{-}5$ -- Apr 9 of $T{-}4$ (2016--17) & Train & \checkmark & \checkmark & \checkmark & \checkmark & ---        \\
Apr 10 of $T{-}4$ -- Apr 9 of $T{-}3$ (2017--18) & Train & \checkmark & \checkmark & \checkmark & ---        & ---        \\
Apr 10 of $T{-}3$ -- Apr 9 of $T{-}2$ (2018--19) & Train & \checkmark & \checkmark & ---        & ---        & ---        \\
Apr 10 of $T{-}2$ -- Apr 9 of $T{-}1$ (2019--20) & Train & \checkmark & ---        & ---        & ---        & ---        \\
\bottomrule
\end{tabular}
\end{table}

We train our model to predict impact scores at five time horizons jointly:
$n \in \{1, 2, 3, 4, 5\}$ years post-publication. This horizon range
balances two competing pressures. Shorter horizons resolve quickly,
allowing recent papers that are distributionally closer to the test set
to contribute training signal. Longer horizons offer more stable measures
of long-term scientific influence, which is the quantity we ultimately
want to predict. Training across horizons jointly extracts signal from
both regimes from a single model.

Label availability for a paper depends on its publication date relative
to the snapshot at which the model is trained. For a model trained at
snapshot date $T$, an $n$-year label for a paper published in year $y$
is observable only if $y + n \leq T$. Older papers contribute supervision
at all five horizons; more recent papers contribute only at the subset
of horizons whose $n$-year window has elapsed by $T$.

Concretely, we partition the dataset as follows. The base training
window covers papers published in $[T-8, T-6]$, for which all five
horizon labels are fully observable. The validation set spans January 1
to April 9 of $T-5$, also with complete label coverage. The augmented
training set covers papers published between April 10 of $T-5$ and
April 9 of $T-1$; each paper in this window contributes supervision
only at the horizons for which its label is observable by the snapshot.
\autoref{tab:multihorizon_split} summarizes per-window label availability
for our headline backtest configuration ($T = 2021$). The April 9
cutoff matches the date of our final data snapshot (April 9, 2026);
restricting validation and test windows to this date ensures consistent
label-coverage across configurations. The test set consists of papers
published between January 1 and April 9 of $T$, with labels assessed
retrospectively against the April 2026 graph snapshot.

\subsection{Architecture and Training}

\definecolor{datafill}{RGB}{241, 245, 249}      
\definecolor{databorder}{RGB}{71, 85, 105}      
\definecolor{opborder}{RGB}{100, 116, 139}      
\definecolor{ffaccent}{RGB}{37, 99, 235}        
\definecolor{fffill}{RGB}{239, 246, 255}        
\definecolor{layerborder}{RGB}{147, 197, 253}   
\definecolor{labelcol}{RGB}{100, 116, 139}      
\definecolor{arrowcol}{RGB}{71, 85, 105}        

\tikzset{
  datablock/.style={
    rectangle, rounded corners=3pt,
    minimum width=2.4cm, minimum height=0.85cm,
    text centered, font=\small\sffamily,
    draw=databorder, fill=datafill,
    line width=0.7pt
  },
  opblock/.style={
    rectangle, rounded corners=3pt,
    minimum width=1.7cm, minimum height=0.85cm,
    text centered, font=\small\sffamily,
    draw=opborder, fill=white,
    line width=0.7pt
  },
  layer/.style={
    rectangle, rounded corners=2pt,
    minimum width=3.4cm, minimum height=0.55cm,
    text centered, font=\scriptsize\sffamily,
    draw=layerborder, fill=white,
    line width=0.5pt
  },
  arrow/.style={-{Stealth[length=5pt]}, line width=0.85pt, color=arrowcol},
  concat/.style={
    circle, minimum size=0.55cm,
    draw=arrowcol, fill=white,
    font=\small\sffamily\bfseries,
    line width=0.7pt
  },
  dimlabel/.style={font=\scriptsize\sffamily\itshape, text=labelcol},
  innerarrow/.style={-{Stealth[length=4pt]}, line width=0.6pt, color=layerborder!70!black},
}

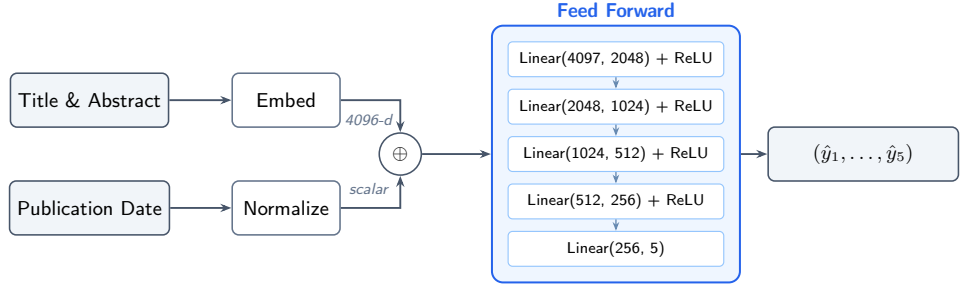
\begin{figure}
  \centering
  \resizebox{0.9\textwidth}{!}{%
      \begin{tikzpicture}
    
        \node[datablock] (text) at (0,  0.85) {Title \& Abstract};
        \node[datablock] (date) at (0, -0.85) {Publication Date};
    
        \node[opblock] (embed) at (3.1,  0.85) {Embed};
        \node[opblock] (norm)  at (3.1, -0.85) {Normalize};
    
        \draw[arrow] (text) -- (embed);
        \draw[arrow] (date) -- (norm);
    
        \node[concat] (cat) at (4.9, 0) {$\oplus$};
        \draw[arrow] (embed.east) -| (cat.north);
        \draw[arrow] (norm.east)  -| (cat.south);
    
        \node[dimlabel] at (4.4,  0.55) {4096-d};
        \node[dimlabel] at (4.4, -0.55) {scalar};
    
        \node[layer] (l3) at (8.3, 0) {Linear(1024, 512) + ReLU};
        \node[layer, above=0.18cm of l3] (l2) {Linear(2048, 1024) + ReLU};
        \node[layer, above=0.18cm of l2] (l1) {Linear(4097, 2048) + ReLU};
        \node[layer, below=0.18cm of l3] (l4) {Linear(512, 256) + ReLU};
        \node[layer, below=0.18cm of l4] (l5) {Linear(256, 5)};
    
        \draw[innerarrow] (l1) -- (l2);
        \draw[innerarrow] (l2) -- (l3);
        \draw[innerarrow] (l3) -- (l4);
        \draw[innerarrow] (l4) -- (l5);
    
        \begin{scope}[on background layer]
          \node[draw=ffaccent, fill=fffill, rounded corners=5pt,
                fit=(l1)(l2)(l3)(l4)(l5), inner sep=7pt,
                line width=0.9pt,
                label={[font=\small\sffamily\bfseries, text=ffaccent]above:Feed Forward}] (ff) {};
        \end{scope}
    
        \draw[arrow] (cat.east) -- (ff.west);
    
        \node[datablock, minimum width=3cm] (out) at (12.2, 0) {$(\hat{y}_1, \ldots, \hat{y}_5)$};
        \draw[arrow] (ff.east) -- (out);
    
      \end{tikzpicture}%
  }
  \caption{Impact prediction model architecture. The title and abstract are encoded by a frozen text embedder (\texttt{llama-embed-nemotron-8b}) into a 4096-dimensional vector, and the publication date is normalized to a scalar. The two are concatenated and passed through a five-layer feed-forward network whose final linear layer produces $\hat{y}_1, \ldots, \hat{y}_5$, the predicted citation impact scores for the first five years following publication.}
  \label{fig:model-architecture}
\end{figure}

The impact prediction model is a feedforward neural network (FFNN) with four hidden layers of dimensions 2048, 1024, 512, and 256, each followed by layer normalization, ReLU activation, and dropout rate 0.3. The network takes the concatenated feature vector as input and produces a 5-vector of impact scores for 1-5 year horizons. All linear layers are initialized using Xavier uniform initialization.

The training loss is the mean of per-horizon MSE losses:
\[
\mathcal{L} = \frac{1}{5} \sum_{n=1}^{5} \mathcal{L}_n,
\]
where $\mathcal{L}_n$ is the MSE between predicted and ground-truth
$n$-year impact scores, computed only over papers in the batch for which
the $n$-year label is observable. Averaging over horizons, rather than
over all valid (paper, horizon) pairs, ensures the 5-year loss term
receives equal weight to the 1-year term despite fewer papers
contributing to it.

Model selection is performed on the 5-year validation Spearman
correlation $\rho$ rather than validation MSE, as our use case
prioritizes ranking papers by predicted impact over predicting their
absolute scores. The checkpoint achieving the highest $\rho$ across the
100 training epochs is retained as the final model. 

We optimize the model using AdamW \cite{DBLP:journals/corr/abs-1711-05101}
with a learning rate of $3 \times 10^{-4}$, weight decay of $10^{-4}$, and
batch size 2048. A linear learning rate warmup is applied over the first
2 epochs, after which the learning rate is held constant. Gradient norms
are clipped to 1.0. Training runs for 100 epochs and completes in
approximately 5 minutes on a single NVIDIA RTX A6000 GPU.

\section{Results}

\subsection{Performance Across Test Years and Horizons}

\begin{figure}
    \centering
    \includegraphics[width=0.8\textwidth]{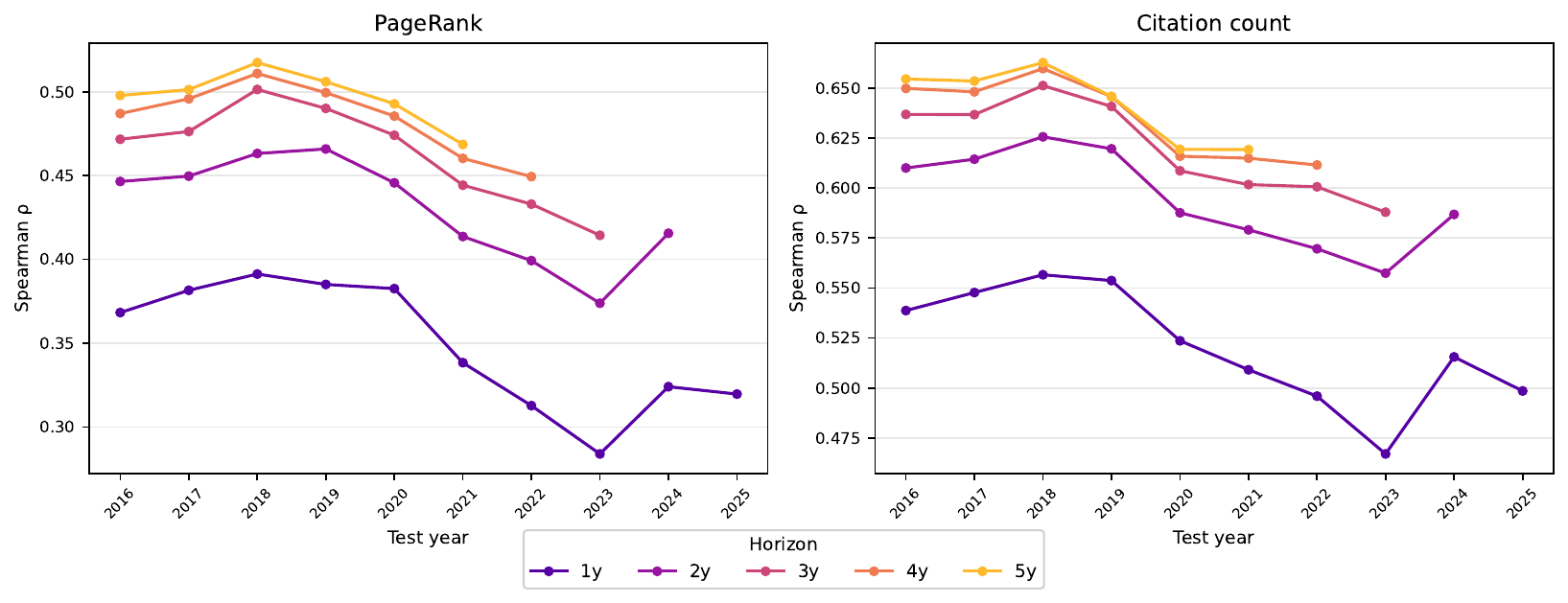}
    \caption{Performance as measuerd by Spearman's $\rho$ for both impact targets across different test years and time horizons.}
    \label{fig:main_performance}
\end{figure}

We evaluate MIRAI across a range of test years for both impact targets. \autoref{fig:main_performance} reports model
performance as measured by Spearman's $\rho$ for both metrics across all
five horizons. At the 5-year horizon and our most recent fully-resolved
test year ($T = 2021$), the citation model achieves $\rho = 0.6192$ and
the PageRank model achieves $\rho = 0.4686$. Note that $n$-year results
are available only for test years $T \le 2026 - n$, since labels for
later test years are not yet observable.

Three trends are immediately visible. First, the citation count model substantially
outperforms the PageRank model across all horizons and test years. We
attribute this to the local-versus-global nature of the two targets:
citation count depends only on which subsequent papers cite the target,
while PageRank depends recursively on the centrality of those citing
papers, which in turn depends on the citing patterns of future work.
The latter is much less inferable from a paper's text alone at
submission time.

Second, performance increases monotonically with inference horizon for both targets. This reflects a difference in label quality rather than
task difficulty: shorter-horizon labels are noisier because papers have
not yet had time to accumulate citations, receive community reaction,
or be built upon by subsequent work, while longer-horizon labels capture
a more stable signal. The benefit of more stable labels evidently
outweighs the additional difficulty of predicting further into the
future.

Third, performance is broadly stable through 2019 but declines from
2020--2023 across both tasks and all horizons. We attribute this to a
widening distributional mismatch between training and test sets, driven
by the rapid expansion of computer science submissions on arXiv
(\autoref{fig:papers_per_year}). Per-field analysis in Appendix~\ref{app:per-field}
supports this account: model performance on computer science papers is
remarkably stable across test years, while the aggregate decline is
driven by physics and mathematics, whose share of arXiv submissions has
shrunk substantially over the same period.

\subsection{Identification of High-Impact Research}

\begin{figure}
    \centering
    \includegraphics[width=0.8\textwidth]{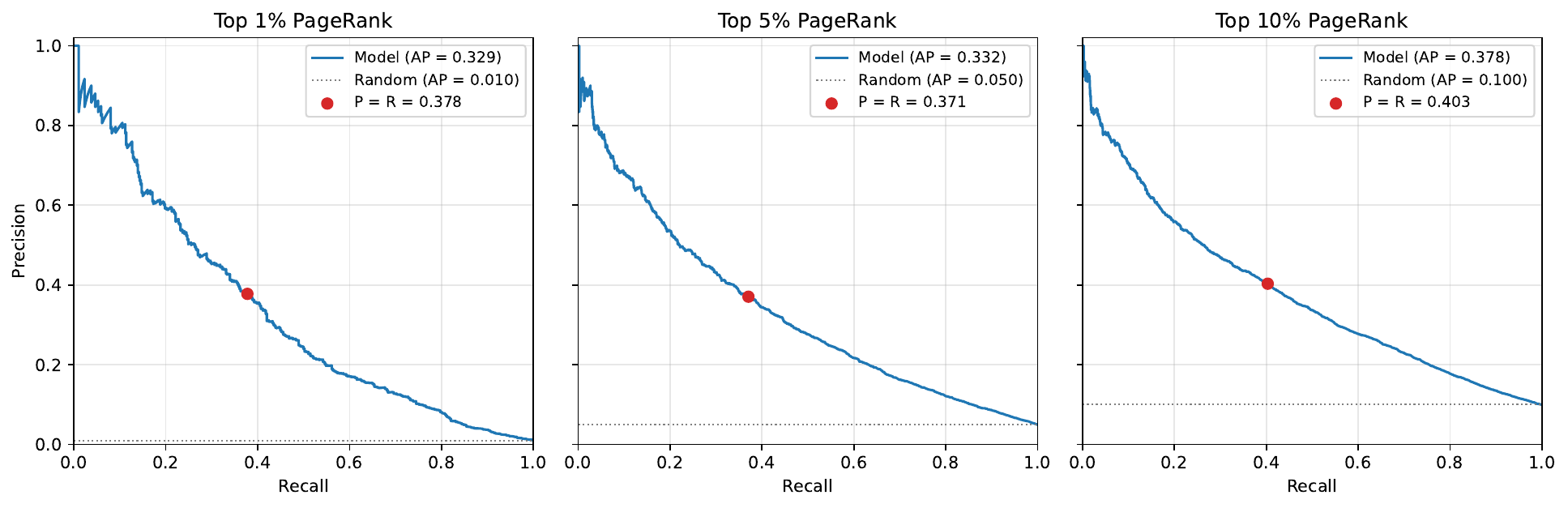}
    \includegraphics[width=0.8\textwidth]{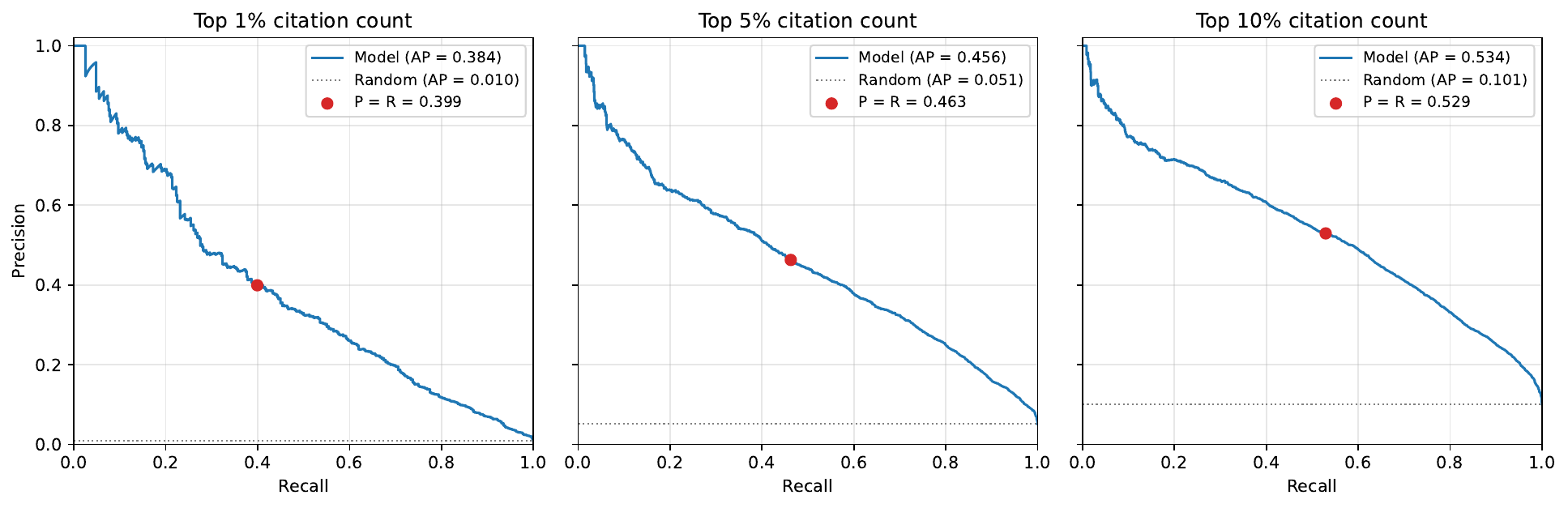}
    \caption{Precision-recall curves for identifying high-impact papers using the PageRank (top) and citation count (bottom) models. High-impact papers are defined using the top 1\%, 5\%, and 10\% of the true impact distribution.}
    \label{fig:precision_recall}
\end{figure}

While global ranking metrics such as Spearman's $\rho$ summarize agreement over the full ranking, they give substantial weight to the relative ordering of low-impact papers. However, our primary goal is to surface papers that are likely to become highly influential. We therefore evaluate the models as high-impact retrieval systems. Specifically, we define papers in the top $x$\% of the true impact distribution as high-impact, with $x \in \{1\%, 5\%, 10\%\}$. \autoref{fig:precision_recall} shows precision-recall curves for these three cutoffs. Across thresholds, both model types exhibit substantial lift over the random baseline, indicating that the models concentrate high-impact papers near the top of their rankings.

\subsection{MIRAI Outperforms Zero-Shot LLMs}

We compare MIRAI against zero-shot LLM baselines. We select LLMs whose knowledge cutoffs are old enough to not include a test set of papers, while remaining recent enough to be representative of frontier model quality. We select two models to satisfy this constraint: GPT-4o \cite{Hurst2024GPT4oSC}, with a knowledge cutoff in October 2023, and Hermes 3 \cite{Teknium2024Hermes3T}, a full-parameter fine-tune of LLaMA-3.1 405B with a cutoff of December 31, 2023. To make the comparison temporally consistent, we restrict our trained citation model to information available through December 31, 2023. We therefore compare all models on their ability to rank papers published between January 1, 2024 and April 9, 2024 by future 2-year citation impact, yielding a test set of 62{,}555 papers.

The LLM baselines are prompted using only the paper title and abstract. Each LLM is queried with temperature 0 and prompted to reason about the paper's novelty, technical contribution, and likely community reception before producing a single non-negative integer corresponding to its estimate of future citation impact. This integer is then used directly as the LLM baseline score for ranking papers. The full prompt is provided in Appendix~\ref{app:llm-prompts}.

\begin{table}
\centering
\caption{Comparison between MIRAI, GPT-4o, and Hermes 3 across global ranking, top-weighted ranking, high-impact retrieval, and cumulative impact concentration metrics when trying to predict 2-year citation counts.}
\label{tab:llm_baseline_comparison}
\begin{tabular}{lccccc}
\toprule
Model & Spearman's $\rho$ & AP top 5\% & Gain@5\% \\
\midrule
\textbf{MIRAI} & \textbf{0.581} & \textbf{0.426} & \textbf{36.7\%} \\
GPT-4o & 0.336 & 0.179 & 23.4\% \\
Hermes 3 & 0.305 & 0.151 & 18.8\% \\
\bottomrule
\end{tabular}
\end{table}

\autoref{tab:llm_baseline_comparison} shows that MIRAI substantially outperforms both zero-shot LLM baselines across all evaluated metrics. MIRAI achieves a Spearman's $\rho$ of 0.581, compared with 0.336 for GPT-4o and 0.305 for Hermes 3, indicating MIRAI produces a substantially better global ordering of papers by future impact. MIRAI also performs better in high-impact retrieval, achieving an average precision of 0.426 under the top-5\% high-impact definition, compared with 0.179 for GPT-4o and 0.151 for Hermes 3. Likewise, the top 5\% of papers ranked by MIRAI capture 36.7\% of total future citation impact, compared with 23.4\% for GPT-4o and 18.8\% for Hermes 3. These results indicate that zero-shot LLM prompting provides a meaningful baseline, but that supervised training on the citation prediction task yields substantially stronger ranking performance.





\subsection{Prospective Study}

We identify papers published between April 10, 2025 and April 9, 2026 that are predicted to achieve the highest five-year impact scores according to MIRAI trained on the most recent graph. Appendix~\ref{app:top} reports the twenty highest-ranked papers under each target definition. Because the relevant five-year outcomes are not yet observable, verification of these predictions is left to future retrospective studies.

\section{Research Generation Pipeline}

\subsection{Design}
\label{sec:design}

\autoref{fig:pipeline} shows the design of our research ideation pipeline. The pipeline is built on the intuition that impactful research most often emerges from combining an established line of work with a recent development.

Each trial pairs an \emph{old} paper (drawn from the top 1\% by ground-truth PageRank within a field, from 2016--2021) with a \emph{new} paper (drawn from the top 5\% by MIRAI-predicted PageRank in the same field, from 2026). Both papers come from the same field of study to ensure productive synthesis. The pair is fed to an LLM generator (Claude Haiku 4.5; the full prompt appears in Appendix~\ref{app:llm-prompts}), which synthesizes the two papers' key contributions into a candidate title and abstract. This is repeated $N$ times to produce $N$ candidate proposals; MIRAI then scores each candidate, and the highest-scoring one is emitted as the pipeline's output. We use MIRAI trained up to April 9, 2026, and fix its prediction date to April 10, 2026 for generated papers. 

We use pairs of seed papers because synthesis between two existing works is a more cognitively realistic model of research ideation than extension of a single paper, and produces a more diverse candidate distribution. A single seed tends to constrain the generator toward incremental extensions of that seed; three or more papers tends to over-constrain generation toward survey-flavored outputs that aggregate existing literature rather than propose new directions.

We use Claude Haiku 4.5 as the generator because it is a frontier-class model at a price point compatible with the scale of generation our experiments require. We choose $N = 32$ to balance two failure modes: too few candidates, and best-of-$N$ selection provides little improvement over a single draw; too many, and the selection step becomes vulnerable to reward hacking, where MIRAI's preferred candidate exploits superficial embedding features rather than reflecting genuine impact.

\definecolor{datafill}{RGB}{241, 245, 249}      
\definecolor{databorder}{RGB}{71, 85, 105}      
\definecolor{opborder}{RGB}{100, 116, 139}      
\definecolor{ffaccent}{RGB}{37, 99, 235}        
\definecolor{fffill}{RGB}{239, 246, 255}        
\definecolor{labelcol}{RGB}{100, 116, 139}      
\definecolor{arrowcol}{RGB}{71, 85, 105}        

\tikzset{
  datablock/.style={
    rectangle, rounded corners=3pt,
    minimum width=2.4cm, minimum height=0.95cm,
    text centered, font=\small\sffamily, align=center,
    draw=databorder, fill=datafill, line width=0.7pt
  },
  opblock/.style={
    rectangle, rounded corners=3pt,
    minimum width=2.0cm, minimum height=0.95cm,
    text centered, font=\small\sffamily, align=center,
    draw=opborder, fill=white, line width=0.7pt
  },
  factorblock/.style={
    rectangle, rounded corners=3pt,
    minimum width=2.0cm, minimum height=0.95cm,
    text centered, font=\small\sffamily, align=center,
    draw=ffaccent, fill=fffill, text=ffaccent, line width=0.9pt
  },
  arrow/.style={-{Stealth[length=5pt]}, line width=0.85pt, color=arrowcol},
  concat/.style={
    circle, minimum size=0.55cm, draw=arrowcol, fill=white,
    font=\small\sffamily\bfseries, line width=0.7pt
  },
  fdash/.style={dashed, draw=ffaccent, line width=0.5pt},
}

\usetikzlibrary{positioning, shapes, arrows.meta, calc}

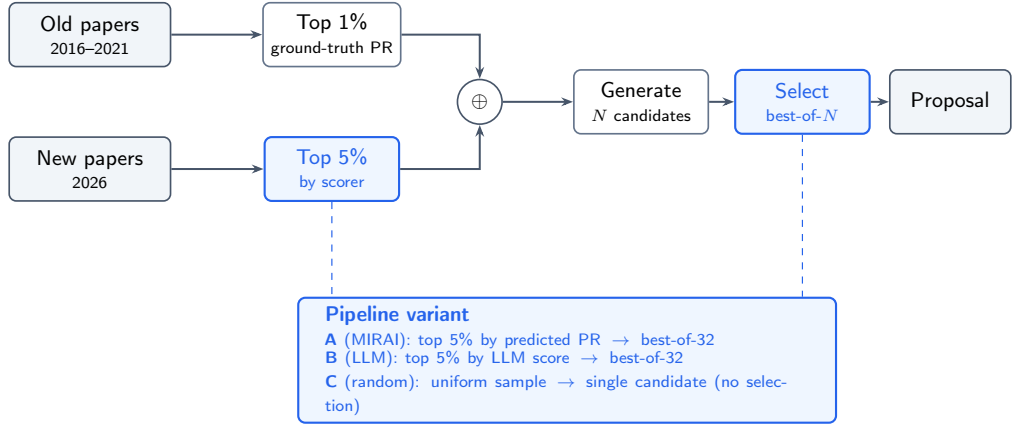
\begin{figure}
\centering
\resizebox{0.95\textwidth}{!}{%
\begin{tikzpicture}

  \node[datablock] (oldcorpus) at (0,  1.0) {Old papers \\ {\scriptsize 2016--2021}};
  \node[datablock] (newcorpus) at (0, -1.0) {New papers \\ {\scriptsize 2026}};

  \node[opblock]     (oldfilt) at (3.6,  1.0) {Top 1\% \\ {\scriptsize ground-truth PR}};
  \node[factorblock] (newfilt) at (3.6, -1.0) {Top 5\% \\ {\scriptsize by scorer}};

  \draw[arrow] (oldcorpus) -- (oldfilt);
  \draw[arrow] (newcorpus) -- (newfilt);

  \node[concat] (cat) at (5.8, 0) {$\oplus$};
  \draw[arrow] (oldfilt.east) -| (cat.north);
  \draw[arrow] (newfilt.east) -| (cat.south);

  \node[opblock]     (gen) at (8.2, 0)  {Generate \\ {\scriptsize $N$ candidates}};
  \node[factorblock] (sel) at (10.6, 0) {Select \\ {\scriptsize best-of-$N$}};
  \node[datablock, minimum width=1.8cm] (out) at (12.8, 0) {Proposal};

  \draw[arrow] (cat.east) -- (gen.west);
  \draw[arrow] (gen) -- (sel);
  \draw[arrow] (sel) -- (out);

  \node[factorblock, align=left, minimum width=80mm, text width=72mm,
        minimum height=18mm, anchor=north]
        (variant) at (7.1, -2.9)
        {\textbf{Pipeline variant} \\[2pt]
         {\scriptsize
          \textbf{A} (MIRAI): top 5\% by predicted PR $\,\to\,$ best-of-32 \\
          \textbf{B} (LLM): top 5\% by LLM score $\,\to\,$ best-of-32 \\
          \textbf{C} (random): uniform sample $\,\to\,$ single candidate (no selection)}};

  \draw[fdash] (newfilt.south) -- (variant.north -| newfilt);
  \draw[fdash] (sel.south)     -- (variant.north -| sel);

\end{tikzpicture}%
}
\caption{Research ideation pipeline. Highlighted (blue) stages depend on the pipeline variant, which determines both the new-paper filter and the candidate selector together; the old-paper filter, generator, and output structure are held constant across all three variants.}
\label{fig:pipeline}
\end{figure}

\subsection{Evaluation}

We evaluate the pipeline described in \ref{sec:design}---variant A in what follows---by comparing it against two alternative configurations that replace its MIRAI-dependent stages. Variant B substitutes Claude Haiku 4.5 for MIRAI at both filtering stages: it scores recent papers to identify the top 5\% pool (full prompt in Appendix~\ref{app:llm-prompts}) and then selects the highest-ranked candidate from the 32 generations. Variant C removes curation entirely, sampling a new paper uniformly within field and generating a single candidate without best-of-$N$ selection. The old paper, generator, and synthesis prompt are held constant across all three variants; only the highlighted stages of \autoref{fig:pipeline} differ.

Each trial generates one candidate per variant, with all three variants sharing the same old paper. The three candidates are then evaluated by an LLM-as-judge from a different model family than the generator (GPT-5.4; full prompt in Appendix~\ref{app:llm-prompts}) on the three pairwise comparisons A vs.\ B, A vs.\ C, and B vs.\ C. We use pairwise rather than absolute ratings because LLM-judges produce more reliable preference signal from direct comparison than from numeric scoring on the same content \cite{liusie2024llmcomparativeassessmentzeroshot}. The judge rates each proposal on four criteria---novelty, technical soundness, significance, and clarity of contribution---on a 1--10 scale and then declares an overall winner (A, B, or tie). To control for first-position bias \cite{Zheng2023JudgingLW}, each comparison is presented in both A/B orderings, with the judge blind to which proposal came from which variant; verdicts that flip across orderings are coded as ties. The judge runs at temperature 0 to minimize stochastic variance; residual provider-side nondeterminism is not characterized. We run 300 trials.

As a secondary analysis, we additionally use MIRAI itself as a judge on the same three pairwise comparisons. For each pair, both candidates are scored by MIRAI and the higher-scoring candidate wins. While we expect variant A to be preferred over variant C by construction, we use this analysis to investigate how abstracts produced by variant B are scored by MIRAI.

\subsection{Results}

\begin{table}
\centering
\footnotesize
\caption{Pairwise win rates from both judges across 300 trials. Y and X denote the first and second pipelines named in each comparison. Each cell reports the proportion of trials with the given verdict and a 95\% Wilson confidence interval. MIRAI is continuous-valued, so its judgments have no ties.}
\label{tab:results-pairwise}
\setlength{\tabcolsep}{4pt}
\begin{tabular}{l ccc ccc}
\toprule
& \multicolumn{3}{c}{\textbf{LLM judge}} & \multicolumn{3}{c}{\textbf{MIRAI judge}} \\
\cmidrule(lr){2-4} \cmidrule(lr){5-7}
Comparison & Y wins & X wins & Tie & Y wins & X wins & Tie \\
\midrule
A vs.\ B & 0.220 {\tiny [0.18, 0.27]} & $\mathbf{0.620}$ {\tiny [0.56, 0.67]} & 0.160 {\tiny [0.12, 0.21]} & $\mathbf{0.963}$ {\tiny [0.94, 0.98]} & 0.037 {\tiny [0.02, 0.07]} & --- \\
A vs.\ C & $\mathbf{0.483}$ {\tiny [0.43, 0.54]} & 0.360 {\tiny [0.31, 0.42]} & 0.157 {\tiny [0.12, 0.20]} & $\mathbf{1.000}$ {\tiny [0.99, 1.00]} & 0.000 {\tiny [0.00, 0.01]} & --- \\
B vs.\ C & $\mathbf{0.697}$ {\tiny [0.64, 0.75]} & 0.160 {\tiny [0.12, 0.21]} & 0.143 {\tiny [0.11, 0.19]} & $\mathbf{0.790}$ {\tiny [0.74, 0.83]} & 0.210 {\tiny [0.17, 0.26]} & --- \\
\bottomrule
\end{tabular}
\end{table}

\begin{table}
\centering
\footnotesize
\caption{Per-criterion mean score differences (Y$-$X) from the LLM judge, on a 1--10 scale, with 95\% bootstrap confidence intervals (10,000 iterations). Bold entries have CIs that exclude zero.}
\label{tab:results-rubric}
\setlength{\tabcolsep}{6pt}
\begin{tabular}{l rrrr}
\toprule
Comparison & Novelty & Technical & Significance & Clarity \\
\midrule
A vs.\ B & $\mathbf{-0.63}$ {\tiny [$-0.82,-0.45$]} & $\mathbf{-1.17}$ {\tiny [$-1.39,-0.95$]} & $\mathbf{-0.40}$ {\tiny [$-0.56,-0.25$]} & $\mathbf{-0.60}$ {\tiny [$-0.79,-0.42$]} \\
A vs.\ C & $+0.01$ {\tiny [$-0.19,+0.21$]} & $-0.06$ {\tiny [$-0.29,+0.18$]} & $\mathbf{+0.82}$ {\tiny [$+0.66,+0.99$]} & $\mathbf{+0.24}$ {\tiny [$+0.03,+0.45$]} \\
B vs.\ C & $\mathbf{+0.69}$ {\tiny [$+0.51,+0.89$]} & $\mathbf{+1.10}$ {\tiny [$+0.88,+1.31$]} & $\mathbf{+1.22}$ {\tiny [$+1.07,+1.37$]} & $\mathbf{+0.86}$ {\tiny [$+0.66,+1.06$]} \\
\bottomrule
\end{tabular}
\end{table}

Pairwise win rates from both judges are reported in \autoref{tab:results-pairwise}. The LLM judge prefers Pipeline B (LLM-curated) over Pipeline A (MIRAI-curated) in 62.0\% of trials, with Pipeline A winning 22.0\% and 16.0\% tied. Pipeline B also substantially outperforms Pipeline C, the no-curation baseline. Pipeline A's advantage over Pipeline C is smaller but still statistically significant.

MIRAI as judge produces a different ordering. Pipeline A wins 96.3\% of A-vs-B comparisons and 100\% of A-vs-C comparisons---an expected pattern given that A's outputs were chosen to maximize the MIRAI score. The informative result here is B-vs-C: MIRAI rates Pipeline B's outputs above Pipeline C's in 79.0\% of trials, despite playing no role in either side's selection.

\autoref{tab:results-rubric} decomposes each LLM-judge comparison into per-criterion mean score differences ($Y - X$) on the 1--10 scale of the rubric. The A-vs-B gap is consistent across all four criteria but largest on technical soundness (mean difference $-1.17$, favoring B). The A-vs-C comparison is more uneven: A and C are statistically indistinguishable on novelty and technical soundness (both 95\% CIs include zero), but A substantively exceeds C on significance ($+0.82$) and modestly on clarity ($+0.24$). B-vs-C shows B preferred on every criterion, with the largest gaps on significance ($+1.22$) and technical soundness ($+1.10$).

The two judges agree that curation improves over the no-curation baseline---both prefer Pipeline B over Pipeline C decisively. They disagree on which form of curation produces better outputs: the LLM judge prefers LLM-curated over MIRAI-curated outputs by 62.0\% to 22.0\%, while MIRAI prefers MIRAI-curated outputs by 96.3\% to 3.7\%.

\subsection{Discussion}

The experiment yields two main findings. First, both judges agree that selecting among candidates substantially improves quality over not selecting: Pipeline B beats Pipeline C in 69.7\% of LLM-judged trials and 79.0\% of MIRAI-judged trials. The agreement is meaningful because Pipeline B's selector and the MIRAI judge use entirely different signals yet arrive at compatible verdicts when comparing curated to uncurated outputs.

Second, the two judges sharply disagree about which form of curation produces better outputs. The LLM judge prefers Pipeline B (LLM-curated) over Pipeline A (MIRAI-curated) at roughly a 3:1 ratio, with B winning across all four rubric criteria. MIRAI prefers Pipeline A over Pipeline B more decisively. MIRAI's near-unanimous preference for Pipeline A is largely tautological---A's outputs were chosen to maximize MIRAI's score among 32 candidates, so MIRAI necessarily ranks them above an unbiased sample. While the LLM judge's preference for Pipeline B is more consequential since the judge and selector drawn from different model families, the two components still share aesthetic preferences common to large language models trained on similar corpora with similar RLHF objectives.

\paragraph{Alignment between MIRAI and LLM scoring.}
The judge disagreement on A versus B raises a central question: to what extent does MIRAI's scoring align with how a language model assesses research quality? The per-criterion breakdown in \autoref{tab:results-rubric} is informative here. MIRAI selection (Pipeline A versus C) leaves novelty and technical soundness statistically unchanged but produces meaningful gains in significance ($+0.82$) and clarity ($+0.24$); LLM selection (Pipeline B versus C) produces large positive gains on all four criteria. MIRAI thus appears to capture a real but partial signal---something that correlates with rubric assessments of significance and clarity, but not with novelty or technical soundness as the LLM judges them, a result that is consistent with MIRAI's training objective.

\paragraph{Implications for the generation pipeline.}
The results suggest that, under rubric-based evaluation, an LLM-based scorer is a more effective candidate selector than MIRAI in the present pipeline. This may not necessarily generalize to long-term citation outcome, and tracking the generated abstracts forward in time to compare citation accrual is a natural follow-up. The practical implication is that MIRAI-based selection improves over no selection but does not match LLM-based selection by rubric standards; a hybrid selector combining the two scores would likely outperform either alone.

\paragraph{Limitations.}
Several factors limit the strength of these conclusions. The LLM judge is a single model (GPT-5.4), and prior work has documented systematic biases in LLM-as-judge evaluations, including verbosity preferences and self-preference within model families \cite{Zheng2023JudgingLW, panickssery2024llmevaluators}. We mitigated position bias by running each comparison in both orderings, but a single judge model cannot be ruled out as a source of systematic preference. The trial set covers a single recent publication window in three fields (Computer Science, Physics, Mathematics), and results may differ in other domains or time periods. Finally, pairwise rubric judging is itself a proxy for paper impact rather than a direct measurement; agreement between two LLM-based components on a rubric does not establish that their preference reflects what produces genuinely impactful research.

Four representative generated research ideas from the MIRAI-based pipeline (variant A) are reproduced in Appendix~\ref{app:generated}.

\section{Conclusion}

We have presented MIRAI, a deep learning framework for predicting paper-level impact across multiple time horizons from a paper's title, abstract, and publication date, along with a research ideation pipeline that uses MIRAI's signal to generate novel research proposals. MIRAI achieves Spearman's $\rho$ of 0.62 on 5-year citation count and 0.47 on 5-year PageRank for papers published in 2021, substantially outperforming zero-shot frontier LLMs (GPT-4o, Hermes 3) on both global ranking and high-impact retrieval. Across test years and horizons, performance increases monotonically with horizon length and declines for recent test years, consistent with longer horizons producing more stable labels and a widening train-test distributional gap driven by accelerating arXiv growth. In a three-variant study of the generation pipeline, any form of candidate curation substantially outperforms an uncurated baseline, though LLM-based and MIRAI-based curation are preferred by different judges, suggesting they capture overlapping but distinct dimensions of paper quality. In conclusion, MIRAI represents a step toward scalable, content-driven assessment of scientific impact and toward AI-assisted ideation, offering a path to both navigating and contributing to a rapidly growing scientific literature.

\begin{ack}
We thank Network Computing Systems (NeCSys) at the MIT Media Lab for providing computing resources and Green Sands Equity for funding this research.
\end{ack}

\bibliographystyle{unsrtnat}
{\small\bibliography{references}}


\newpage
\appendix

\section{Per-Field Performance Results}
\label{app:per-field}

\begin{figure}[htbp]
    \centering
    \includegraphics[width=0.8\linewidth]{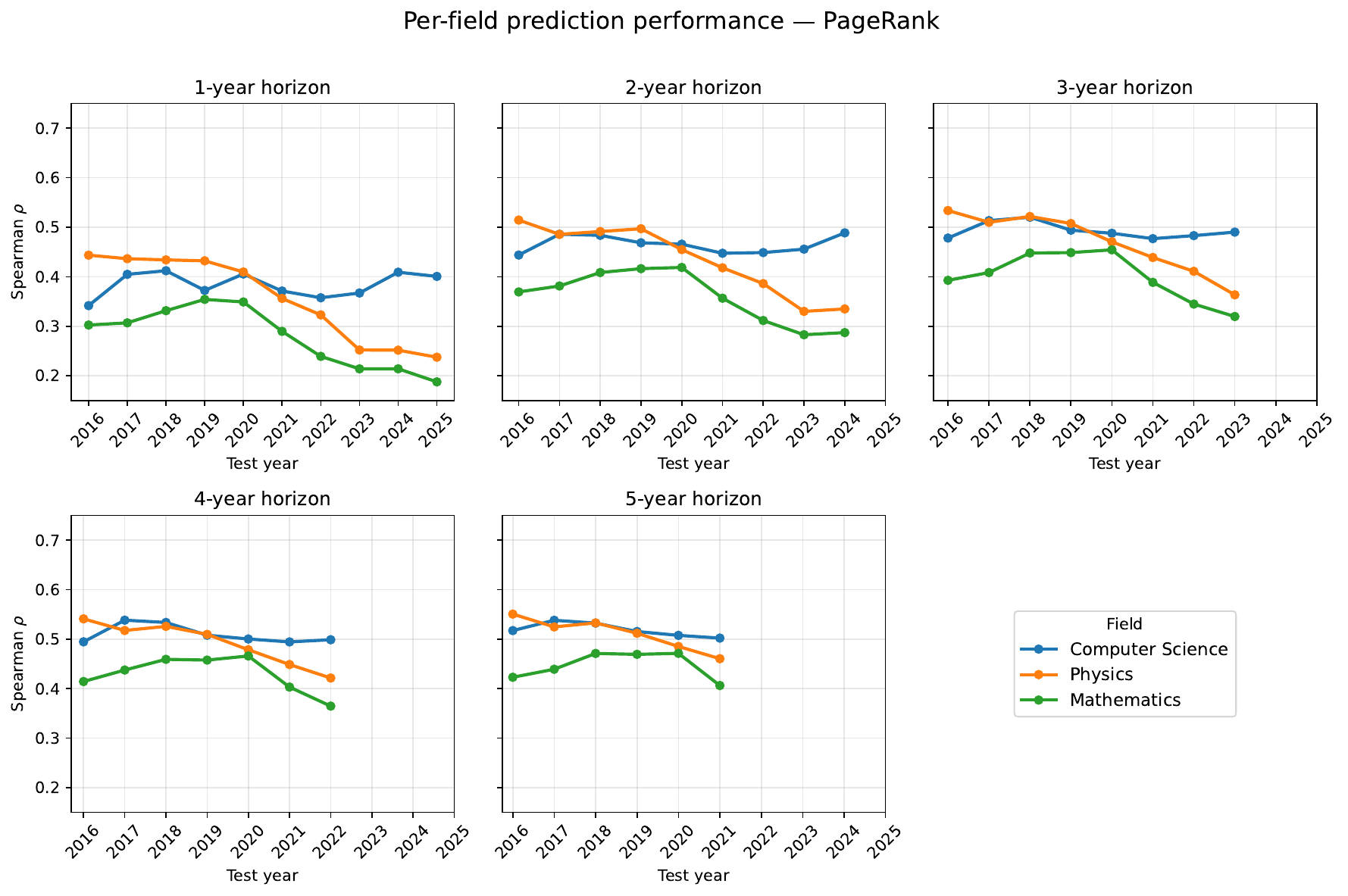}
    \includegraphics[width=0.8\linewidth]{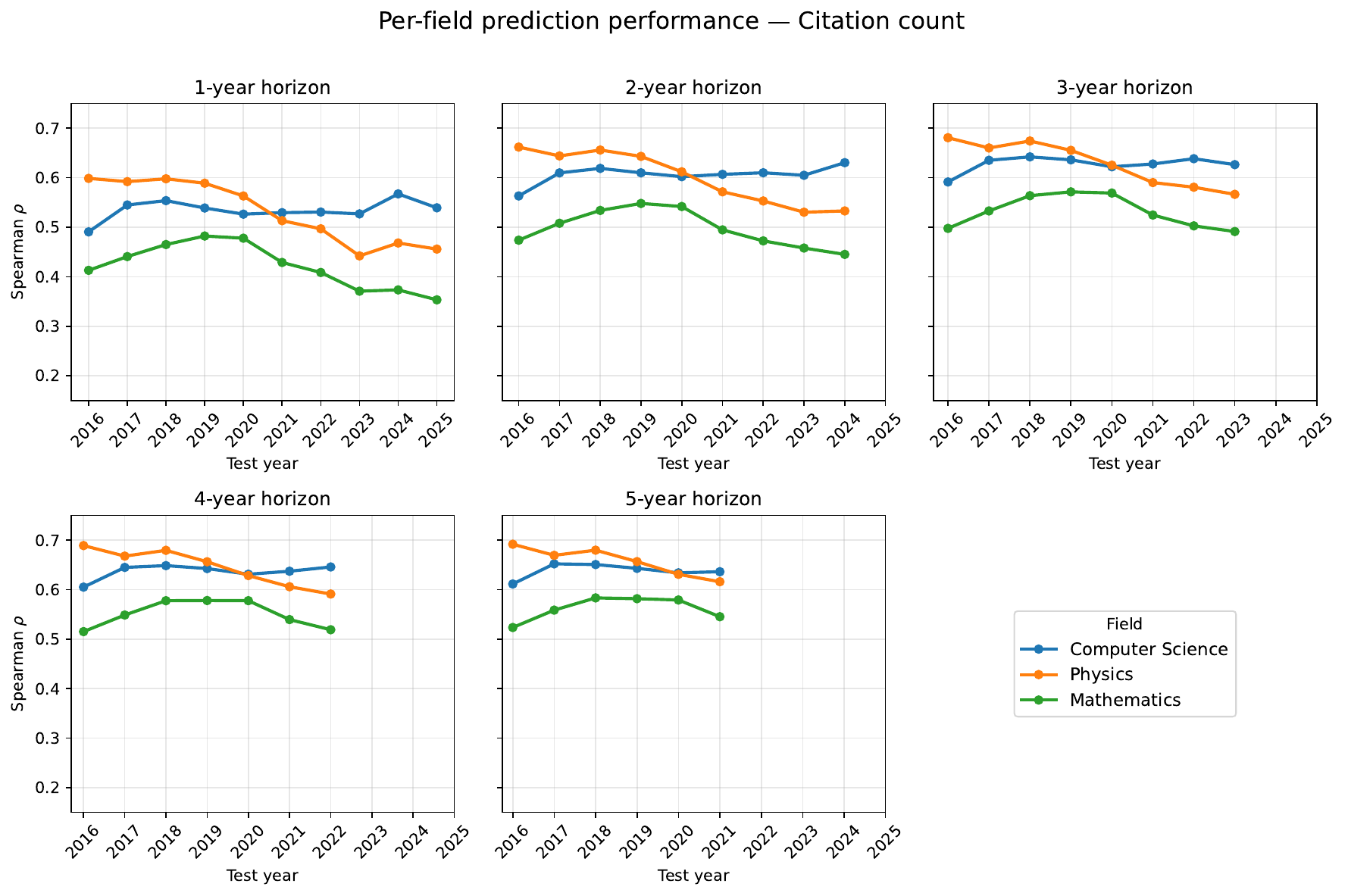}
    \caption{Per-field performance plots for PageRank (top) and citation (bottom) models.}
    \label{fig:placeholder}
\end{figure}

\section{LLM Prompts}
\label{app:llm-prompts}

\subsection{LLM baseline for citation prediction}

\textbf{System prompt}
\begin{verbatim}
Predict a calibrated citation count. Return exactly one non-negative 
integer.
\end{verbatim}

\textbf{User prompt}
\begin{verbatim}
You are a bibliometrics baseline model.

Goal:
Predict the number of citations this paper will have 24 months after 
publication.

You only have the title and abstract. Make a best-effort prediction.

Internally estimate:
1. likely research field
2. breadth of audience
3. novelty
4. methodological importance
5. practical usefulness
6. whether this sounds incremental or field-shaping

Then output your best estimate as one non-negative integer.

Do not output a default value.
Do not choose a number merely because it is common.
Do not explain.
Do not output JSON.
Output only one integer.

Title:
{title}

Abstract:
{abstract}
\end{verbatim}

\subsection{LLM scoring to select top 5\% research as generation seeds}

\begin{verbatim}
You are evaluating recent academic papers for likely scholarly impact. By
"impact" we mean the likelihood that the paper, if published, would receive
substantial citations within 3-5 years and meaningfully advance its field.

Below are papers from the field of {field}. Score each paper on a 1.0-10.0
scale:
  - 1.0-3.0: low expected impact (incremental, narrow audience)
  - 4.0-6.0: typical paper
  - 7.0-8.0: above-average expected impact
  - 9.0-10.0: exceptional, potentially field-defining

Calibrate so that ~5 is typical and 9+ is reserved for exceptional work. Be
discriminating; flat scores across all papers are not useful. Do NOT reward
buzzwords, length, polish, or grandiose framing.

Papers:

{papers}

Respond ONLY with valid JSON:
{{
  "scores": {{"<paperId>": <score>, ...}}
}}
Include every paper listed.
\end{verbatim}

\subsection{Novel research generation}

\begin{verbatim}
You are a researcher proposing a new paper. Below are two papers from 
the recent literature. Drawing on them — by extending, combining, 
contrasting, or addressing a gap they reveal — propose a new research 
paper.

Paper 1:
Title: {title_1}
Abstract: {abstract_1}

Paper 2:
Title: {title_2}
Abstract: {abstract_2}

Write a title and abstract for the proposed paper. Follow the 
conventions of a real published abstract: briefly state the motivation, 
describe the approach, summarize the key contribution or result, and 
close with a sentence on implications. Aim for 150-250 words. Write in 
a neutral, declarative tone — describe what the paper does rather than 
asserting that it is novel or important.

Output exactly in this format:

Title: <title>
Abstract: <abstract>
\end{verbatim}

\subsection{LLM selection of best generated candidate}

\begin{verbatim}
You are evaluating proposed research papers for likely scholarly impact. By
"impact" we mean the likelihood that the paper, if published, would receive
substantial citations within 3-5 years and meaningfully advance its field.

Below are {n} proposed research papers, each with a title and abstract.
Identify the single proposal most likely to achieve high scholarly impact.

Consider:
  - Novelty: a genuinely new idea, method, or finding
  - Technical soundness: methodologically rigorous and feasible
  - Significance: matters to a substantial research community
  - Clarity: contribution is precisely articulated

Do NOT reward buzzwords, length, verbosity, or grandiose framing.

Proposed papers:

{candidates}

Respond ONLY with valid JSON:
{{
  "reasoning": "<2-3 sentences justifying the choice>",
  "selected": <integer index, 0-based>
}}
\end{verbatim}

\subsection{LLM judge for comparing generated research}

\begin{verbatim}
You are evaluating two proposed research papers for likely scholarly impact.
Rate each on these criteria, 1-10:

  - novelty: introduces a genuinely new idea, method, or finding
  - technical: methodologically rigorous, feasible, well-grounded
  - significance: matters to a substantial research community
  - clarity: the contribution is precisely articulated

Then choose an overall winner: "A", "B", or "tie" if the two are roughly 
equivalent.

Be discriminating. Do NOT reward buzzwords, length, verbosity, or grandiose 
framing. Score papers on substantive content.

Paper A:
Title: {a_title}
Abstract: {a_abstract}

Paper B:
Title: {b_title}
Abstract: {b_abstract}

Respond ONLY with valid JSON in this schema:
{{
  "A_scores": {{"novelty": <int>, "technical": <int>, "significance": <int>, 
  "clarity": <int>}},
  "B_scores": {{"novelty": <int>, "technical": <int>, "significance": <int>, 
  "clarity": <int>}},
  "reasoning": "<2-3 sentences>",
  "verdict": "A" | "B" | "tie"
}}
\end{verbatim}

\section{Predicted High-Impact Papers}
\label{app:top}

\begin{longtable}{r
    >{\raggedright\arraybackslash}p{0.4\textwidth}
    >{\raggedright\arraybackslash}p{0.4\textwidth}}
\caption{Top 20 predicted high-impact papers by target metric published between April 10, 2025 and April 9, 2026.}
\label{tab:top20-predicted-impact} \\
\toprule
\textbf{Rank} & \textbf{Citation count target} & \textbf{PageRank target} \\
\midrule
\endfirsthead

\toprule
\textbf{Rank} & \textbf{Citation count target} & \textbf{PageRank target} \\
\midrule
\endhead

\midrule
\multicolumn{3}{r}{\emph{Continued on next page}} \\
\endfoot

\bottomrule
\endlastfoot

1 & OpenThoughts: Data Recipes for Reasoning Models
  & Qwen3 Technical Report \\

2 & GLM-4.5V and GLM-4.1V-Thinking: Towards Versatile Multimodal Reasoning with Scalable Reinforcement Learning
  & gpt-oss-120b\&gpt-oss-20b Model Card \\

3 & Audio Flamingo 3: Advancing Audio Intelligence with Fully Open Large Audio Language Models
  & Gemini 2.5: Pushing the Frontier with Advanced Reasoning, Multimodality, Long Context, and Next Generation Agentic Capabilities \\

4 & Qwen3 Technical Report
  & RoboBrain 2.0 Technical Report \\

5 & Ovis2.5 Technical Report
  & Qwen3-Omni Technical Report \\

6 & Seed1.5-VL Technical Report
  & rStar2-Agent: Agentic Reasoning Technical Report \\

7 & Qwen3-Omni Technical Report
  & Qwen3-VL Technical Report \\

8 & gpt-oss-120b\&gpt-oss-20b Model Card
  & Llama-Nemotron: Efficient Reasoning Models \\

9 & RoboBrain 2.0 Technical Report
  & OpenThoughts: Data Recipes for Reasoning Models \\

10 & Qwen3-VL Technical Report
   & DeepMath-103K: A Large-Scale, Challenging, Decontaminated, and Verifiable Mathematical Dataset for Advancing Reasoning \\

11 & Gemini 2.5: Pushing the Frontier with Advanced Reasoning, Multimodality, Long Context, and Next Generation Agentic Capabilities
   & Audio Flamingo 3: Advancing Audio Intelligence with Fully Open Large Audio Language Models \\

12 & rStar2-Agent: Agentic Reasoning Technical Report
   & DeepSeek-V3.2: Pushing the Frontier of Open Large Language Models \\

13 & Emu3.5: Native Multimodal Models are World Learners
   & Large Language Models Hallucination: A Comprehensive Survey \\

14 & InternVL3: Exploring Advanced Training and Test-Time Recipes for Open-Source Multimodal Models
   & INTELLECT-3: Technical Report \\

15 & MiniMax-M1: Scaling Test-Time Compute Efficiently with Lightning Attention
   & Magistral \\

16 & SAIL-VL2 Technical Report
   & Mobile-Agent-v3: Fundamental Agents for GUI Automation \\

17 & AM-Thinking-v1: Advancing the Frontier of Reasoning at 32B Scale
   & Ovis2.5 Technical Report \\

18 & InternVL3.5: Advancing Open-Source Multimodal Models in Versatility, Reasoning, and Efficiency
   & This Time is Different: An Observability Perspective on Time Series Foundation Models \\

19 & Kimi K2: Open Agentic Intelligence
   & Seed1.5-VL Technical Report \\

20 & EO-1: An Open Unified Embodied Foundation Model for General Robot Control
   & GLM-4.5V and GLM-4.1V-Thinking: Towards Versatile Multimodal Reasoning with Scalable Reinforcement Learning \\

\end{longtable}

\section{Sampled Generated Paper Ideas}
\label{app:generated}

\begin{longtable}{p{0.95\textwidth}}
\caption{Four representative Pipeline A (MIRAI-curated) selections.}\label{tab:samples-pipeline-a} \\
\toprule
\endfirsthead

\multicolumn{1}{c}{\textit{Table~\ref{tab:samples-pipeline-a} continued from previous page}} \\
\toprule
\endhead

\midrule
\multicolumn{1}{r}{\textit{continued on next page}} \\
\endfoot

\bottomrule
\endlastfoot

\textbf{Forward Model Optimization for Quantitative Ultrasound Imaging: A Systematic Mathematical Framework and Open-Source Implementation} \\*
\small Quantitative ultrasound imaging techniques such as computed ultrasound tomography in echo mode (CUTE) rely on accurate forward models to reconstruct tissue properties from phase-shifted echo measurements. While recent advances have improved forward model accuracy through empirical refinements, a systematic mathematical-physical optimization framework for these models remains underdeveloped. This work presents a comprehensive approach to forward model design for speed-of-sound imaging that combines rigorous mathematical optimization with practical computational efficiency. We develop an open-source toolbox that implements vector space decomposition methods to characterize the relationship between spatial sound speed distributions and measured phase shifts, explicitly accounting for transmit-receive angle centering and echo localization errors. The framework enables both direct forward mapping and iterative optimization of model parameters. We validate the approach in phantom studies mimicking liver tissue and demonstrate that systematic optimization yields substantially improved quantitative reconstruction compared to empirically-refined models. Additionally, we show that the optimized forward model generalizes across different transducer geometries and imaging scenarios. The toolbox is made available to the research community to standardize forward model development and enable broader exploration of design alternatives. These results suggest that systematic mathematical optimization of forward models can enhance the diagnostic accuracy and robustness of quantitative ultrasound imaging systems for clinical translation. \\
\midrule
\textbf{Interpretable Feature Representations for Long-Horizon Molecular Dynamics Trajectories} \\*
\small Generative models for protein dynamics simulation have achieved impressive performance on short-horizon predictions but struggle to maintain physical plausibility over microsecond timescales. A critical bottleneck lies in the lack of interpretable feature representations that capture the essential dynamical characteristics of molecular trajectories, making it difficult to diagnose failure modes, compare models systematically, and transfer knowledge across simulation tasks. We propose applying canonical time-series characteristics to molecular dynamics, extracting a minimal set of interpretable features from protein trajectories that reflect their physical and dynamical properties. Drawing on the catch22 framework—which identifies essential time-series features across diverse domains—we develop MD-catch22, a curated feature set specifically tailored to protein dynamics that includes metrics for conformational stability, inter-atomic correlation structure, energy landscape properties, and temporal coherence. We evaluate MD-catch22 on trajectory data from 50 protein systems of varying complexity, demonstrating that these 22 features can effectively discriminate between high-fidelity and degraded trajectories while remaining computationally efficient to compute. We further show that models can be compared and ranked using these features with 89\% agreement to expensive physical validation metrics. Our approach provides a common interpretable language for assessing and improving long-horizon molecular dynamics models, enabling systematic model development and broader accessibility to trajectory-quality evaluation across computational biology. \\
\midrule
\textbf{Generative Design of Metasurface-Based Optical Systems: Bridging Semantic Intent and Electromagnetic Performance} \\*
\small The design of advanced optical systems increasingly relies on metasurfaces—subwavelength-engineered surfaces capable of arbitrary wavefront manipulation—yet their synthesis remains confined to specialized expertise. While recent advances have demonstrated both high-performance metasurface implementations and generative design frameworks for conventional refractive optics, a unified approach to metasurface design from functional specifications remains absent. This work presents an end-to-end generative framework that extends semantic-to-physical translation methodology to metasurface-based systems by integrating large language models with a differentiable electromagnetic simulation engine. The framework autonomously interprets user specifications and generates metasurface architectures optimized for anomalous reflection, refraction, and polarization control across specified frequency ranges and angular response requirements. We demonstrate the approach on three representative applications: (1) wide-angle beam steering metasurfaces for radar and communication systems, (2) achromatic metalenses spanning visible and infrared bands, and (3) multifunctional metasurface arrays supporting simultaneous control of multiple electromagnetic properties. Each design is validated against full-wave electromagnetic simulations and fabrication constraints. By combining the semantic reasoning capabilities of LLMs with the physical accuracy of electromagnetic solvers, this framework addresses the gap between intuitive design intent and the complex optimization landscape of metasurface engineering. The results establish metasurface design as an accessible domain for non-specialists while maintaining performance standards required for practical deployment. \\
\midrule
\textbf{Escaping Hallucination Saddle Points: A Trajectory-Based Optimization Framework for Deep Research Agents} \\*
\small Deep Research Agents (DRAs) frequently encounter failure modes characterized by stable but suboptimal reasoning trajectories—what we term "hallucination saddle points"—where agents become trapped in plausible but incorrect research directions. These intermediate failures, including flawed planning and biased summarization, accumulate throughout the research process, yet existing optimization approaches treat them as endpoint failures. We propose a trajectory-aware optimization framework that applies second-order optimization principles to navigate the hallucination landscape of DRAs. Our key insight is that hallucination manifolds exhibit saddle-point-like geometry: locally stable under first-order metrics (e.g., superficial coherence) but unstable under second-order properties (e.g., fact consistency and logical grounding). We extend perturbed gradient descent to the discrete decision space of research agents, introducing stochastic perturbations to the planning module that efficiently escape these hallucination saddle points. Using the PIES taxonomy to characterize hallucination geometry, we show that trajectory-aware second-order optimization achieves convergence to higher-quality research outputs with sample complexity that scales poly-logarithmically in trajectory length. Experiments on DeepHalluBench demonstrate that agents augmented with our framework reduce hallucination propagation by 35-48\% while maintaining computational efficiency. This work bridges non-convex optimization theory and LLM agent reliability, offering principled methods for diagnosing and escaping systematic failure modes in research trajectories. \\
\end{longtable}


\end{document}